\documentclass[12pt,a4]{article}
\usepackage{amsmath}
\usepackage{bm}
\usepackage{amsfonts}
\usepackage{amssymb}
\usepackage{graphics}
\usepackage[normal]{caption2}
\usepackage{subfigure}
\usepackage{rotating}
\usepackage{citesort}
\def\be{\begin{equation}}
\def\ee{\end{equation}}
\def\ba{\begin{array}}
\def\ea{\end{array}}
\def\bea{\begin{equation}}
\def\bea{\end{equation}}
\def\bea{\begin{eqnarray}}
\def\eea{\end{eqnarray}}

\begin{document}
\baselineskip 20pt \setlength\tabcolsep{2.5mm}
\renewcommand\arraystretch{1.5}
\setlength{\abovecaptionskip}{0.1cm}
\setlength{\belowcaptionskip}{0.5cm}
\pagestyle{empty}
\newpage
\pagestyle{plain} \setcounter{page}{1} \setcounter{lofdepth}{2}
\begin{center} {\large\bf Role of different model ingredients in the exotic cluster-decay of $^{56}$Ni$^*$}\\
\vspace*{0.4cm}
{\bf Narinder K. Dhiman}\footnote{Email:~narinder.dhiman@gmail.com}\\
{\it  Govt. Sr. Sec. School, Summer Hill, Shimla -171005, India}
\end{center}
We present cluster decay studies of $^{56}$Ni$^*$ formed in
heavy-ion collisions using different Fermi density and nuclear
radius parameters proposed by various authors. Our study reveals
that different technical parameters do not alter the transfer
structure of fractional yields significantly. The cluster decay
half-lives of different clusters lies within $\pm$10\% for
different Fermi density parameters and nuclear radius, therefore,
justify the current set of parameters used in the literature for
the calculations of cluster decay.


\newpage
\baselineskip 20pt
\section{Introduction}
    In earlier days, nucleus was considered to have a
uniform density and sharp radius. With the passage of time, the
density distribution was found to be more complicated. Several
different forms (direct or indirect) exist in literature that can
explain these complicated nuclear density distributions. The first
method is the direct parametrization which involves the choice of
a suitable functional form where parameters are varied to fit the
experimental data. The two parameter Fermi density distribution is
an example of such a parametrization. The second method is of
indirect parametrization of density distribution proceeds via
nuclear models. The nuclear models like shell model contains
certain parameters which are determined by other physical
considerations and it is then used to calculate the nuclear
density distribution without further adjustments. The experimental
data can be described accurately with two-parameter Fermi density
distribution at relatively low momentum. Among all the density
distributions two-parameters Fermi density has been quite
successful in the low, medium and heavy mass regions. The
systematic study of charge distributions have been carried out in
Refs.\cite{Angeli,Weso,Fried}. We shall use this density
distribution here.
\par

Since the nuclear systems obey quantum laws, therefore, their
surfaces are not well defined. The nuclear density remains
constant up to certain distance but fall more rapidly close to the
surface region where the nucleons are free to move about. The
nuclear densities provide important information about the
structure of nuclear matter at low energies and other important
information regarding the equation of state at intermediate
energies~\cite{sk,rk}.
\par

Various methods have been developed for exploring the nuclear
structure and radius. The electron scattering/ electrically
charged particles of high energy are employed as probe to explore
the proton distribution of the nuclei (i.e charge radii), whereas
neutral nuclear probes such as neutrons will give the effect of
nuclear forces over the nuclear surface (i.e. interaction radii).
The charge radii are often used to extract the information about
nuclear radii. The electron scattering experiments shows that the
charge distribution within a nucleus either follow Fermi
trapezoidal shape or modified Gaussian distribution. These studies
have shown that nuclear charge density does not decrease abruptly
but has a finite diffuseness.
\par

A model that uses density distribution such as two parameter Fermi
density (as shown in Fig.~1) has to rely on the information about
nuclear radius (or half density radii $R_0$), central density
$\rho_0$, and surface diffuseness ($a$). Interestingly, several
different experimental as well as theoretical values of these
parameters are available in
literature~\cite{8Ngo,8Puri92,8SM,8EW,8HS,Elton}. In addition,
several different names such as central radii, equivalent sharp
radii, root mean square radii etc. have also been used in the
literature to define different functional forms. The role of
different  radii was examined in exotic cluster decay
half-lives~\cite{8Rkg} and interestingly two different forms of
radii were found to predict five order of magnitude different
half-lives within the same theoretical model. Similarly, the use
of different values of surface diffuseness also varies from author
to author. The effect of these model ingredients on the fusion
process at low incident energy have been studied in
Ref.~\cite{8RashmiT} and there was found that the effect of
different radii is more than marginal and therefore this parameter
should be used with a more fundamental basis. Unfortunately, no
systematic study is still available in the cluster decay process.
In this paper, we plan to study the role of Fermi density
parameters in the cluster decay of $^{56}$Ni$^*$ when formed in
heavy-ion collisions. This study is still missing in the
literature.
\par
Heavy-ion reactions provide a very good tool to probe the nucleus
theoretically. This includes low energy fusion process~\cite{id},
intermediate energy phenomena~\cite{qmd} as well as cluster-decay
and/or formation of super heavy nuclei~\cite{gupta,kp}. In the
last one decade, several theoretical models have been employed in
the literature to estimate the half-life times of various exotic
cluster decays of radioactive nuclei. These outcome have also been
compared with experimental data. Among all the models employed
preformed cluster model (PCM)~\cite{rkg88,mal89,kum97} is widely
used to study the exotic cluster decay. In this model the
clusters/ fragments are assumed to be pre-born well before the
penetration of the barrier. This is in contrast to the unified
fission models (UFM)~\cite{7Poen,7Buck,7Sand}, where only barrier
penetration probabilities are taken into account. In either of
these approach, one needs complete knowledge of nuclear radii and
densities in the potential.
\par

Cluster decay of $^{56}$Ni is studied when formed as an excited
compound system in heavy-ion reactions. Since $^{56}$Ni has
negative $Q$-value (or $Q_{out}$) and is stable against both
fission and cluster decay processes. However, if is is produced in
heavy-ion reactions depending on the incident energy and angular
momentum involved, the excited compound system could either
fission, decay via cluster emissions or results in resonance
phenomenon. The $^{56}$Ni has a negative $Q_{out}$ having
different values for various exit channels and hence would decay
only if it were produced with sufficient compound nucleus
excitation energy $E^{\ast}_{CN}~(=E_{cm} + Q_{in})$, to
compensate for negative $Q_{out}$, the deformation energy of the
fragments $E_d$, their total kinetic energy ($TKE$) and the total
excitation energy ($TXE$), in the exit channel as:
\begin{equation}
E^{\ast}_{CN} = \mid Q_{out} \mid + E_{d} + TKE + TXE.
 \label{eq:1}
\end{equation}
(see Fig.~2, where $E_d$ is neglected because the fragments are
considered to be spherical).  Here $Q_{in}$ adds to the entrance
channel kinetic energy $E_{cm}$ of the incoming nuclei in their
ground states.
\par

Section 2 gives some details of the Skyrme energy density model
and preformed cluster model and its simplification to unified
fission model. Our calculations for the decay half-life times of
$^{56}$Ni compound system and a discussion of the results are
presented in Section 3. Finally, the results are summarized in
Section 4.

\section{Model}
\subsection{Skyrme Energy Density Model}

In the Skyrme Energy Density Model (SEDM)~\cite{8Puri92}, the
nuclear potential is calculated as a difference of energy
expectation value $E$ of the colliding nuclei at a finite distance
$R$ and at complete isolation (i.e. at
$\infty$)~\cite{8Puri92,VB72}.
\begin{equation}
V_{N} (R) = E(R)- E(\infty), \label{eq:2}
\end{equation}

where $E = \int H (\vec{r})\vec{dr}$, with $H(\vec{r})$ as the
Skyrme Hamiltonian density which reads as:
\begin{eqnarray}
H(\rho,\tau,\vec{J}) & = & \frac{\hbar^2}{2m} \tau +\frac{1}{2}t_0
[(1+
\frac{1}{2}x_0)\rho^2-(x_0+\frac{1}{2})(\rho_n^2+\rho_p^2)]\nonumber\\
& &+\frac{1}{4}( t_1+t_2)\rho\tau  +\frac{1}{8}(t_2-t_1)(\rho_n
\tau_n+\rho_p \tau_p) \nonumber \\ & & +\frac{1}{16} (t_2-3t_1)
\rho \nabla^2 \rho + \frac{1}{4}t_3 \rho_n \rho_p \rho \nonumber
\\ & &+\frac{1}{32} (3t_1+t_2) (\rho_n \nabla^2 \rho_n+\rho_p
\nabla^2 \rho_p) \nonumber \\ && -\frac {1}{2} W_0(\rho
\vec{\nabla}\cdot\vec{J} +\rho_n \vec{\nabla}\cdot\vec{J}_n+
\rho_p \vec{\nabla} \cdot\vec{J}_p). \label{eq:3}
\end{eqnarray}
Here  $\vec{J} = \vec{J}_{n} + \vec{J}_{p}$ is the spin density
which was generalized by Puri et al.~\cite{8Puri92}, for
spin-unsaturated nuclei and  $\tau=\tau_{n} + \tau_{p}$ is the
kinetic energy density calculated using Thomas Fermi
approximation~\cite{gupta85,2Von}, which reduces the dependence of
energy density $H(\rho ,\tau ,\vec{J})$ to be a function of nucleon
density $\rho$ and spin density $\vec{J}$ only. Here strength of
surface correction factor is taken to be zero (i.e. $\lambda=0$).
The remaining term is the nucleon density $\rho=\rho_{n} + \rho_{p}$
is taken to be well known two-parameter Fermi density. The Coulomb
effects are neglected in the above energy density functional, but
will be added explicitly. In Eq.~(\ref{eq:3}), six parameters $t_0$,
$t_1$, $t_2$, $t_3$, $x_0$, and $W_0$ are fitted by different
authors to obtain the best description of the various ground state
properties for a large number of nuclei. These different
parameterizations have been labeled as S, SI, SII, SIII etc. and
known as Skyrme forces for light and medium colliding nuclei. Other
Skyrme forces are able to reproduce the data for heavy systems
better. The Skyrme force used for the present study is SIII with
parameters as: $t_0=-1128.75$ MeVfm$^3$, $t_1=395.00$ MeVfm$^5$,
$t_2=-95.00$ MeVfm$^5$, $t_3=14000.00$ MeVfm$^6$, $x_0=0.45$, and
$W_0=120.00$ MeVfm$^5$. It has been shown  in previous studies that
SIII force reproduces the fusion barrier much better than other sets
of Skyrme forces for light and medium  nuclei. Other Skyrme forces
such as SKa, SKm, however, are found to be better for heavier
masses.
\par

From Eq.~(\ref{eq:3}), one observes that the Hamiltonian density
$H(\rho ,\tau ,\vec{J})$ can be divided into two parts: (i) the
spin-independent part $V_{P}(R)$, and (ii) spin-dependent
$V_{J}(R)$~\cite{8Puri92} as:
\begin{eqnarray}
V_{N}(R)  &=&\int \left\{H(\rho)- \left[H_{1}(\rho_{1}) +
H_{2}(\rho_{2}) \right] \right\}d\vec{r}
\nonumber \\
&+&\int \left\{H(\rho, \vec{J})- \left[H_{1}(\rho_{1},
\vec{J}_{1})
+ H_{2}(\rho_{2}, \vec{J}_{2}) \right] \right\}d\vec{r}\nonumber \\
&=& V_{P}(R)+V_{J}(R) \label{eq:4}
\end{eqnarray}
We apply the standard Fermi mass density distribution for
nucleonic density:
\begin{equation}
\rho (R)=\frac{\rho_{0} }{ 1+ \exp\left\{\frac{R - R_{0} }{a}
\right\} },~~~~~~~~~~~~ - \infty \leq R \leq \infty \label{eq:5}
\end{equation}
Here $\rho_{0}$, $R_{0}$ and {\it ``a''} are respectively, the
average central density, half-density radius and the surface
diffuseness parameter. The $R_{0}$ gives the distance where density
drops to the half of its maximum value and the surface thickness
$s~(=4.4a)$ has been defined as the distance over which the density
drops from 90\% to 10\% of its maximum value is the average central
density $\rho_{0}$. The systematic two parameter Fermi density
distribution is shown in Fig.~1.
\par

Another quantity, which is equally important is the r.m.s. radius
$\langle r^{2}\rangle _{m}$ defined as:
\begin{eqnarray}
\left<r^{2} \right>_{m} &= & \int r^{2}\rho \left(\vec{r}
\right)d\vec{r}
  = 4\pi \int\limits_{0}^{\infty } \rho \left(\vec{r} \right)r^{4}d^{3}r.
\label{eq:6}
\end{eqnarray}
One can find the half density radius by varying surface
diffuseness {\it ``a''} and keeping r.m.s. radius $\langle
r^{2}\rangle _{m}$ constant or from normalization condition:
\begin{equation}
R_{0}= \frac{1}{3}\left[5 \left<r^{2} \right>_{m} -7\pi ^{2}a^{2}
\right], \label{eq:7}
\end{equation}
The average central density $ \rho_{0}$ given by \cite{8Stancu}
\begin{equation}
\rho_{0} =\frac{3A}{4 \pi R^{3}_{0}}
\left[1+\frac{\pi^{2}a^{2}}{R^{2}_{0}} \right]^{-1}. \label{eq:8}
\end{equation}
Using Eq.~(\ref{eq:5}), one can find the density of neutron and
proton individually as:
\begin{equation}
\rho _{n}= \frac{N}{A}\rho ,~~~~~~~~~~\rho _{p}= \frac{Z}{A}\rho.
\label{eq:9}
\end{equation}
For the details of the model, reader is referred to
Ref.~\cite{8Puri92}.
\par

In order to see the effect of different Fermi density parameters
on the cluster decay half-lives, we choose the following different
Fermi density parameters proposed by various authors.
\begin{enumerate}
    \item[1.] \textbf{H. de Vries  \emph{ et al.}~\cite{Elton}:} Here, we use the interpolated
experimental data~\cite{8Puri91} of Elton and H. de Vries for half
density radius $R_{0}$ and surface thickness $a$. Using $R_{0}$
and $a$, central density $\rho_{0}$ can be computed using
Eq.~(\ref{eq:7}). This set of parameters is labeled as DV.

    \item[2.] \textbf{Ng\^o-Ng\^o~\cite{8Ngo}:} In the version of Ng\^o-Ng\^o, a simple analytical
expression is used for nuclear densities instead of Hartree-Fock
densities. These densities are taken to be of Fermi type and
written as:
\begin{equation}
\rho_{n, p} (R)= \frac{\rho_{n, p}(0)}{1+\exp [(R-C_{n,
p})/0.55]}~, \label{eq:10}
\end{equation}
$\rho_{n, p}(0)$ are then given by:
\begin{equation}
\rho_n(0)=\frac{3}{4\pi}\frac{N}{A}\frac{1}{r^3_{0_{n}}},~~~~~~~\rho_p(0)=\frac{3}{4\pi}\frac{Z}{A}\frac{1}{r^3_{0_{p}}}~.
\label{eq:11}
\end{equation}
where $C$ represents the central radius of the distribution.
\begin{equation}
C= R\left[1-\frac{1}{R^{2}}\right],
 \label{eq:12}
\end{equation}
and
\begin{equation}
R= \frac{NR_{n} +ZR_{p}}{A}. \label{eq:13}
\end{equation}
The sharp radii for proton and neutron are given by,
\begin{equation}
R_{p}= r_{0_{p}}A^{1/3},~~~~~~~~~~~~~R_{n}= r_{0_{n}}A^{1/3},
\label{eq:14}
\end{equation}
with
\begin{equation}
r_{0_{p}}= 1.128~fm,~ r_{0_{n}}= 1.1375 + 1.875\times 10^{-4} A.
\label{eq:15}
\end{equation}
This set of parameters is labeled as Ngo.

    \item[3.] \textbf{S.A. Moszkwski~\cite{8SM}:} The Fermi density parameters
due to Moszkwski has central density $\rho_0 = 0.16$ nucl./fm$^3$,
the surface diffuseness parameters $a$ is equal to 0.50 fm and
radius $R_0 = 1.15A^{1/3}$. This set of parameters is labeled as
SM.

    \item[4.] \textbf{E. Wesolowski~\cite{8EW}:} The expressions for Fermi density
parameters taken by E. Wesolowski reads as: The central density
\begin{equation}
\rho_0 = \left[\frac{4}{3} \pi R^3_0 \left\{1 + \left(\pi
a/R_0\right)^2 \right\} \right]^{-1}. \label{eq:16}
\end{equation}
\par

The surface diffuseness parameters $a$ = 0.39 fm and half density
radius,
\begin{equation}
R_0 = R^{\prime}\left[1 - \left(\frac{b}{R^{\prime}}\right)^2 +
\frac{1}{3}\left( \frac{b}{R^{\prime}}\right)^6 +
\cdot\cdot\cdot\cdot\cdot\right];\label{eq:17}
\end{equation}
with
\begin{equation}
R^{\prime} = \left[1.2 - \frac{0.96}{A^{1/3}}\left(
\frac{N-Z}{A}\right)
\right]A^{1/3},~\mbox{and}~b=\frac{\pi}{\sqrt3}a. \label{eq:18}
\end{equation}
This set of parameters is labeled as EW.

    \item[5.] \textbf{H. Schechter  \emph{et al.}~\cite{8HS}:} The value of Fermi density
parameters taken by H. Schechter  \emph{et al.} can be summarized
as: central density $\rho_0 = 0.212/(1 + 2.66A^{-2/3})$, the
surface diffuseness parameters $a$ is equal to 0.54 fm and radius
$R_0 = 1.04A^{1/3}$  in single folding model for one of the
nucleus. This set of parameters is labeled as HS.
\end{enumerate}

In the spirit of proximity force theorem, the spin independent
potential $V_{P}(R)$ of the two spherical nuclei, with radii C$_1$
and C$_2$ and whose centers are separated by a distance
$R=s+C_1+C_2$ is given by
\begin{equation}
V_{P}(R)  = 2\pi \overline{R} \phi (s), \label{eq:19}
\end{equation}
where
\begin{equation}
\phi (s) =\int \left\{H(\rho)- \left[H_{1}(\rho_{1}) +
H_{2}(\rho_{2}) \right] \right\}dZ, \label{eq:20}
\end{equation}
and
\begin{equation}
\overline{R} =\frac{C_1 C_2}{C_1+C_2}, \label{eq:21}
\end{equation}
with S\"ussmann central radius $C$ given in terms of equivalent
spherical radius $R$ as
\begin{equation}
C =R-\frac{b}{R}. \label{eq:22}
\end{equation}
Here the surface diffuseness $b=1$ fm and nuclear radius $R$ taken
as given by various authors in the
literature~\cite{8Ngo,3Blocki77,AW95,3MS00,Royer,5Bass73,5CW}.

In the original proximity potential~\cite{3Blocki77}, the
equivalent sharp radii used are
\begin{equation}
R= 1.28A^{1/3}- 0.76 + 0.8A^{-1/3}~~~{\rm fm}.\label{eq:23}
\end{equation}
This radius is labeled as R$_{Prox77}$.

In the present work, we also used the nuclear radius due to Aage
Winther, labeled as R$_{AW}$ and read as~\cite{AW95}:
\begin{equation}
R= 1.20A^{1/3}- 0.09~~~{\rm fm}. \label{eq:24}
\end{equation}
The newer version of proximity potential uses a different form of
nuclear radius~\cite{3MS00}
\begin{equation}
R= 1.240A^{1/3}\left[1+1.646A^{-1}- 0.191A_{s} \right]~~~{\rm
fm}.\label{eq:25}
\end{equation}
This radius is labeled as R$_{Prox00}$.

Recently, a  newer form of above Eq.~(\ref{eq:25}) with slightly
different constants is reported ~\cite{Royer}
\begin{equation}
R= 1.2332A^{1/3}+2.8961A^{-2/3}- 0.18688A^{1/3}A_{s} ~~~{\rm
fm},\label{eq:26}
\end{equation}
and is labeled as R$_{Royer}$.

For  Ng\^o and Ng\^o~\cite{8Ngo} nuclear radius, we use Eqs.
(\ref{eq:13})-(\ref{eq:15}) and is labeled as R$_{Ngo}$.

The potential based on the classical analysis of experimental
fusion excitation functions, used the nuclear radius (labeled as
R$_{Bass}$)~\cite{5Bass73} as:
\begin{equation}
R= 1.16A^{1/3} - 1.39A^{-1/3}. \label{eq:27}
\end{equation}
The empirical potential due to Christensen-Winther (CW) uses the
same radius form (Eq.~(\ref{eq:27})) having different constants
(labeled as R$_{CW}$)~\cite{5CW}.
\begin{equation}
R= 1.233A^{1/3} - 0.978A^{-1/3}.
 \label{eq:28}
\end{equation}
\subsection{The Preformed Cluster Model}
For the cluster decay calculations, we use the Preformed Cluster
Model~\cite{rkg88,mal89,kum97}. It is based on the well known
quantum mechanical fragmentation
theory~\cite{7Puri,rkg75,7SNG,7Maruhn74}, developed for the
fission and heavy-ion reactions and used later on for predicting
the exotic cluster decay~\cite{7Sandu80,Rose,rkg94} also. In this
theory, we have two dynamical collective coordinates of mass and
charge asymmetry $\eta=(A_1-A_2)/(A_1+A_2)$ and
$\eta_Z=(Z_1-Z_2)/(Z_1+Z_2)$. The decay half-life $T_{1/2}$ and
decay constant $\lambda$, in decoupled $\eta$- and $R$-motions is
\begin{equation}
\lambda =\frac{\ln 2}{T_{1/2}}= P_{0}\nu _{0}P,
 \label{eq:29}
\end{equation}
where the preformation probability $P_0$ refers to the motion in
$\eta$ and the penetrability $P$ to $R$-motion. The $\nu_0$ is the
assault frequency with which the cluster hits the barrier. Thus,
in contrast to the unified fission
models~\cite{7Poen,7Buck,7Sand}, the two fragments in PCM are
considered to be pre-born at a relative separation co-ordinate $R$
before the penetration of the potential barrier with probability
$P_0$. The preformation probability $P_0$ is given by
\begin{equation} P_{0}(A_{i})=
\mid \psi(\eta,A_i) \mid^{2}\sqrt{B_{\eta \eta }(\eta
)}\left(\frac{4}{A_i}\right),\,\,\,\,\,\,\,\,\,(i=1~{\rm or}~2),
 \label{eq:30}
\end{equation}
with $\psi^{\nu}(\eta),~~ \nu=0,1,2,3,.....$, as the solutions of
stationary Schr\"odinger equation in $\eta$ at fixed $R$,
\begin{equation}
\left[-\frac{\hbar ^{2}}{2\sqrt{B_{\eta\eta }}}\frac{\partial
}{\partial \eta }\frac{1}{\sqrt{B_{\eta \eta }}}\frac{\partial
}{\partial \eta }+V_R(\eta) \right]\psi^{\nu}(\eta)=E^{\nu}
\psi^{\nu}(\eta), \label{eq:31}
\end{equation}
solved at $R=R_a=R_{min}$ at the minimum configuration i.e. $R_a =
R_{min}$ (corresponding to $V_{min}$) with potential at this
$R_a$-value as $V(R_a = R_{min})= \overline{V}_{min}$ (displayed
in Fig.~2).
\par

The temperature effects are also included here in this model
through a Boltzmann-like function as
\begin{equation}
\mid \psi(\eta)\mid^{2}=\sum_{\nu=0}^{\infty }\mid \psi^{\nu}
(\eta) \mid^{2}\exp \left(-\frac{E_{\eta}}{T} \right),
\label{eq:32}
\end{equation}
where the nuclear temperature $T$ (in MeV) is related
approximately to the excitation energy $E^{\ast}_{CN}$, as:
\begin{equation}
E^{\ast}_{CN}=\frac{1}{9}A{T}^2-T, \qquad\qquad {(\rm in~ MeV)}.
\label{eq:33}
\end{equation}
The fragmentation potential (or collective potential energy)
$V_R(\eta)$,  in Eq.~(\ref{eq:31}) is calculated within Strutinsky
re-normalization procedure, as
\begin{equation}
V_R(\eta)=-\sum^2_{i=1} \left[V_{LDM}(A_i,Z_i)+\delta U_i
\exp\left(-\frac{T^2}{T_0^2}\right)\right] +\frac{Z_1 \cdot
Z_2e^2}{R} + V_N(R),
 \label{eq:34}
\end{equation}
where the liquid drop energies ($V_{LDM} = B - \delta U$) with $B$
as theoretical binding energy of M\"oller et al.~\cite{mol95} and
the shell correction $\delta U$ calculated in the asymmetric two
center shell model. The additional attraction due to nuclear
interaction potential $V_N(R)$ is calculated within SEDM potential
using different Fermi density parameters  and nuclear radii as
discussed earlier. The shell corrections are considered to vanish
exponentially for $E^{\ast}_{CN} \ge 60$ MeV, giving $T = 1.5$
MeV. The mass parameter $B_{\eta\eta}$ representing the kinetic
energy part of the Hamiltonian in Eq.~(\ref{eq:31}) are smooth
classical hydrodynamical masses of Kr\"oger and
Scheid~\cite{kro80}.
\par

The WKB action integral was solved for the penetrability
$P$~\cite{rkg94}. For each $\eta$-value, the potential $V(R)$ is
calculated by using SEDM for $R \ge R_{d}$, with $R_{d}=R_{min} +
\Delta R$ and for $R \le R_{d}$, it is parameterized simply as a
polynomial of degree two in $R$:
\begin{equation}
V(R)=\left \{
\begin{array}{ll}
\mid Q_{out} \mid +{a_1}(R-R_0)+{a_2}(R-R_0)^2 & \mbox{for \quad $R_0\leq R\leq R_{d} $}, \\
V_N(R) + Z_1 \cdot Z_2 e^2/R & \mbox{for $ \quad R\geq R_{d}$},
\end{array}
\right. \label{eq:35}
\end{equation}
where $R_0$ is the parent nucleus radius and $\Delta R$ is chosen
for smooth matching between the real potential and the
parameterized potential (with second-order polynomial in $R$). A
typical scattering potential, calculated by using
Eq.~(\ref{eq:35}) is shown in Fig.~2, with tunneling paths and the
characteristic quantities also marked. Here, we choose the first
(inner) turning point $R_a$ at the minimum configuration i.e. $R_a
= R_{min}$ (corresponding to $V_{min}$) with potential at this
$R_a$-value as $V(R_a = R_{min})= \overline{V}_{min}$  and the
outer turning point $R_b$ to give the $Q_{eff}$-value of the
reaction  $V(R_b) = Q_{eff}$. This means that the penetrability
$P$ with the de-excitation probability, $W_i=\exp (-bE_i)$ taken
as unity, can be written as $P=P_iP_b,$ where $P_i$ and $P_b$ are
calculated by using WKB approximation, as:
\begin{equation}
P_i=\exp \left[- \frac{2}{\hbar} \int\limits_{R_a}^{R_i}\{2\mu
[V(R)-V(R_i)]\}^{1/2}dR \right],
 \label{eq:36}
\end{equation}
and
\begin{equation}
P_b=\exp\left[- \frac{2}{\hbar} \int\limits_{R_i}^{R_b}\{2\mu
[V(R)-Q_{eff}]\}^{1/2}dR \right],
 \label{eq:37}
\end{equation}
here $R_a$ and $R_b$ are, respectively, the first and second
turning points. This means that the tunneling begins at $R =
R_a~(=R_{min})$ and terminates at $R = R_b$, with $V(R_b) =
Q_{eff}$. The integrals of Eqs.~(\ref{eq:36}) and~(\ref{eq:37})
are solved analytically by parameterizing the above calculated
potential $V(R)$.
\par
The assault frequency $\nu_0$ in Eq.~(\ref{eq:29}) is given simply
as
\begin{equation}
\nu_0 = \frac{v}{R_0} = \frac{(2E_2/\mu)^{1/2}}{R_0},
\label{eq:38}
\end{equation}
where $E_2 = \frac{A_1}{A} Q_{eff}$ is the kinetic energy of the
emitted cluster, with $Q_{eff}$ shared between the two fragments
and $\mu =m(\frac{A_1 A_2}{A_1+ A_2})$ is the reduced mass.
\par

The PCM can be simplified to UFM, if preformation probability
$P_0=1$ and the penetration path is straight to $Q_{eff}$-value.

\section{\label{result}Results and Discussions}
In the following, we see the effect of different Fermi density
parameters and nuclear radii on the cluster-decay process using
the Skyrme energy density formalism within PCM and UFM.

First of all, to see the effect of different Fermi density
parameters on the cluster decay half-lives, we choose the
different Fermi density parameters proposed by various authors as
discussed earlier.
\par

Fig.~2 shows the characteristic scattering potential for the
cluster decay of $^{56}$Ni$^{\ast}$ into $^{16}$O + $^{40}$Ca
channel as an illustrative example.  In the exit channel for the
compound nucleus to decay, the compound nucleus excitation energy
$E_{CN}^{\ast}$ goes in compensating the negative $Q_{out}$, the
total excitation energy $TXE$ and total kinetic energy $TKE$ of
the two outgoing fragments as the effective Q-value (i.e.
$TKE=Q_{eff}$ in the cluster decay process). In addition, we plot
the penetration paths for PCM and UFM using Skyrme force SIII
(without surface correction factor, $\lambda =0$) with DV Fermi
density parameters. For PCM, we begin the penetration path at $R_a
= R_{min}$ with potential at this $R_a$-value as $V(R_a =
R_{min})= \overline{V}_{min}$ and ends at $R = R_b$, corresponding
to $V(R=R_b) = Q_{eff}$, whereas for UFM, we begin at $R_a$ and
end at $R_b$ both corresponding to $V(R_a) =V(R_b)=Q_{eff}$. We
have chosen only the case of variable $Q_{eff}$ (as taken in
Ref.~\cite{nkdid}), for different cluster decay products to
satisfy the arbitrarily chosen relation $Q_{eff}=0.4(28 - \mid
Q_{out} \mid)$ MeV, as it is more realistic~\cite{MKS00}. The
scattering potential with SM Fermi density parameters is also
plotted for comparison.
\par

Fig.~3(a) and (b) shows the fragmentation potential $V(\eta)$ and
fractional yield at $R = R_{min}$ with $V(R_{min})=
\overline{V}_{min}$. The fractional yields are calculated within
PCM at $T$ = 3.0 MeV using various Fermi density parameters for
$^{56}$Ni$^{\ast}$. From figure, we observe that different
parameters have minimal role in the fractional mass distribution
yield. The fine structure is not at all disturbed for different
sets of Fermi density parameters.
\par

We have also calculated the half-life times (or decay constants)
of $^{56}$Ni$^{\ast}$ within PCM and UFM for clusters $\ge
^{16}$O. For $^{16}$O, the cluster decay constant varies by an
order of magnitude ten. The variation is much more with SM
parameters. In the case of UFM, variation is almost constant.
\par

In Fig.~4, we display the cluster decay half-lives $\log T_{1/2}$
for various Fermi density parameters using PCM. There is smooth
variation in half-life times with all the density parameters
except for SM parameter. The trends in the variation of cluster
half-life times (or decay constants) are similar in both PCM and
UFM, but in case of UFM  decay constants are more by an order of
ten. In SM the decay constants are larger by an order of 14.

In order to quantify the results, we have also calculated the
percentage variation in $\log T_{1/2}$ as:
\begin{equation}
\left[\log T_{1/2} \right] \% = \frac{(\log T_{1/2})^i- (\log
T_{1/2})^{DV}}{(\log T_{1/2})^{DV}}\times 100, \label{eq:39}
\end{equation}
where $i$ stands for the half-life times calculated using
different Fermi density parameters. The variation in the cluster
decay half-lives is studied with respect to DV parameters. In
Fig.~5(a) and (b), we display the percentage variation in the
half-life times within both the PCM and UFM models as a function
of cluster mass $A_2$ using Eq.~(\ref{eq:39}). For the PCM these
variation lies within $\pm$5\% excluding SM parameters, whereas
including SM parameters it lies within $\pm$13\%. In the case of
UFM half-lives lies within $\pm$1.5\% for all density parameters
except of SM. For SM parameters variations lie within $\pm$9\%.

Finally, it would be of interest to see how different forms of
nuclear radii as discussed earlier would affect the cluter decay
half-lives.

In Fig.~6, we display the characteristic scattering potential for
the cluster decay of $^{56}$Ni$^{\ast}$ into $^{28}$Si + $^{28}$Si
channel for R$_{Bass}$ and R$_{Royer}$ forms of nuclear radius. In
the exit channel for the compound nucleus to decay, the compound
nucleus excitation energy $E_{CN}^{\ast}$ goes in compensating the
negative $Q_{out}$, the total excitation energy $TXE$ and total
kinetic energy $TKE$ of the two outgoing fragments as the
effective Q-value. We plot the penetration path for PCM using
Skyrme force SIII (without surface correction factor, $\lambda
=0$) with nuclear radius R$_{Bass}$. Here again, we begin the
penetration path at $R_a = R_{min}$ with potential at this
$R_a$-value as $V(R_a = R_{min})= \overline{V}_{min}$ and ends at
$R = R_b$, corresponding to $V(R=R_b) = Q_{eff}$  for PCM.  The
$Q_{eff}$ are same as discussed earlier.

Fig.~7(a) and (b) show the fragmentation potentials $V(\eta)$ and
fractional yields at $R = R_{min}$ with $V(R_{min})=
\overline{V}_{min}$. The fractional yields are calculated within
PCM at $T$ = 3.0 MeV for $^{56}$Ni$^{\ast}$ using various forms of
nuclear radii. From figure, we observe that different radii gives
approximately similar behavior, however small changes in the
fractional mass distribution yields are observed. The fine
structure is not at all disturbed for different radius values.
\par

We have also calculated the half-life times (or decay constants)
of $^{56}$Ni$^{\ast}$ within PCM  for clusters $\ge^{16}$O. The
cluster decay constant for nuclear radius due to Bass varies by an
order of magnitude $10^{2}$, where as order of magnitude is same
for other radii. In Fig.~8, we display the cluster decay
half-lives $\log T_{1/2}$ for various nuclear radii taken by
different authors as explained earlier using PCM. One can observe
small variations in half-life times.

In order to quantify the results, we have also calculated the
percentage variation in $\log T_{1/2}$ as:
\begin{equation}
\left[\log T_{1/2} \right] \% = \frac{(\log T_{1/2})^i- (\log
T_{1/2})^{R_{Royer}}}{(\log T_{1/2})^{R_{Royer}}}\times 100,
\label{eq:40}
\end{equation}
where $i$ stands for the half-life times calculated using
different forms of nuclear radii. The variation in the cluster
decay half-lives is studied with respect to radius formula given
by Royer R$_{Royer}$. In Fig.~9, we display the percentage
variation in the half-life times for PCM as a function of cluster
mass $A_2$ using Eq.~(\ref{eq:40}). These variation lies within
$\pm$7\% excluding Bass radius where it lies within $\pm$10\%.

\section{\label{summary}Summary}

We here  reported the role of various model ingredients as well as
radii in the cluster decay constant calculations. Our studies
revealed that the effect of different density and nuclear radius
parameters on the cluster decay half-life times is about $10\%$.
Our study justify the use of current set of parameters for radius as the effect of different prescriptions is very small.\\


\newpage

\begin{figure}[!t]
\centering \vskip 1cm
\includegraphics[angle=0,width=14cm]{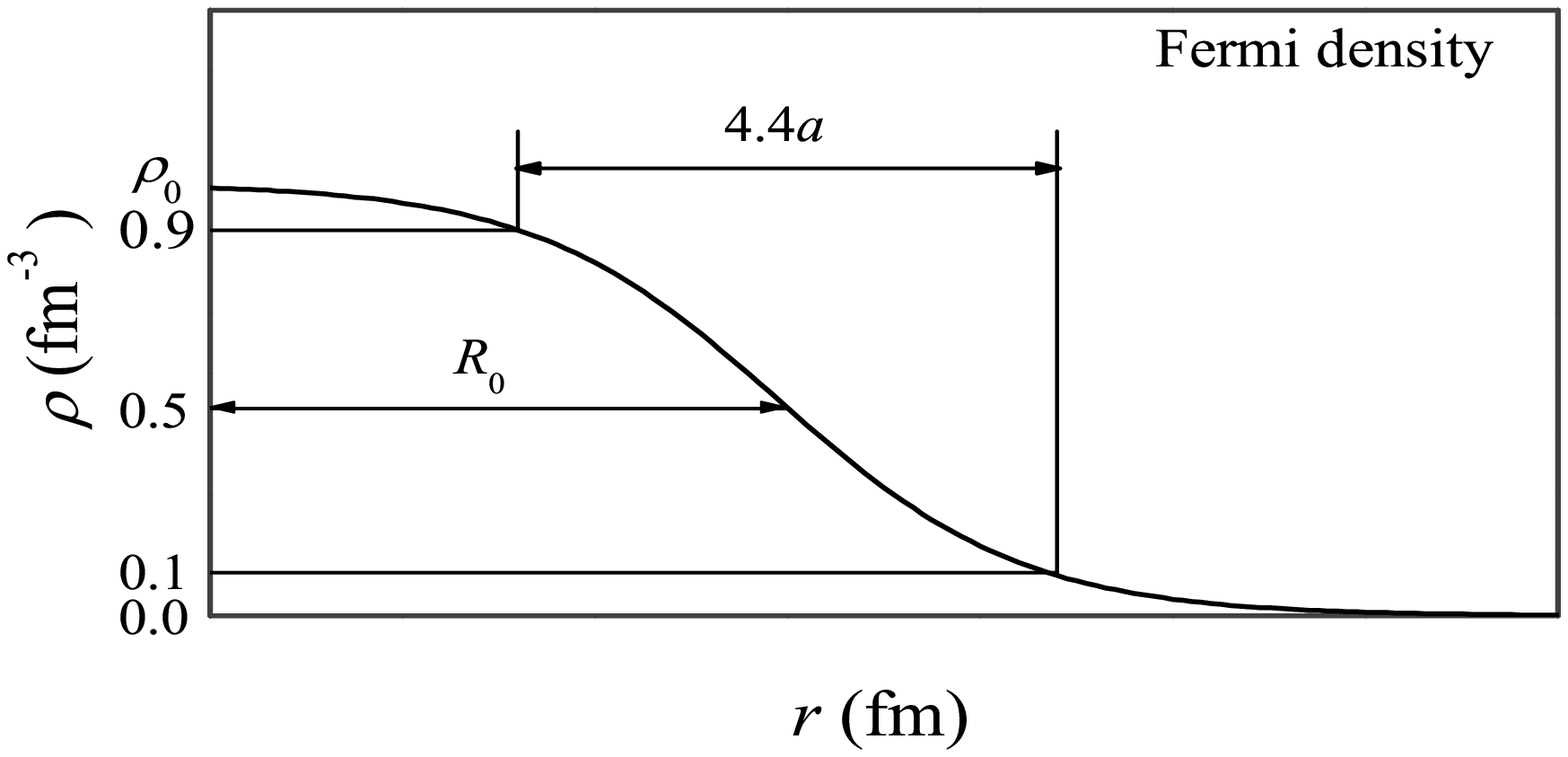}
\vskip -4cm \caption{The systematics diagram for two parameter
Fermi density.}\label{fig2}
\end{figure}

\begin{figure}[!t]
\centering \vskip 1cm
\includegraphics[angle=0,width=14cm]{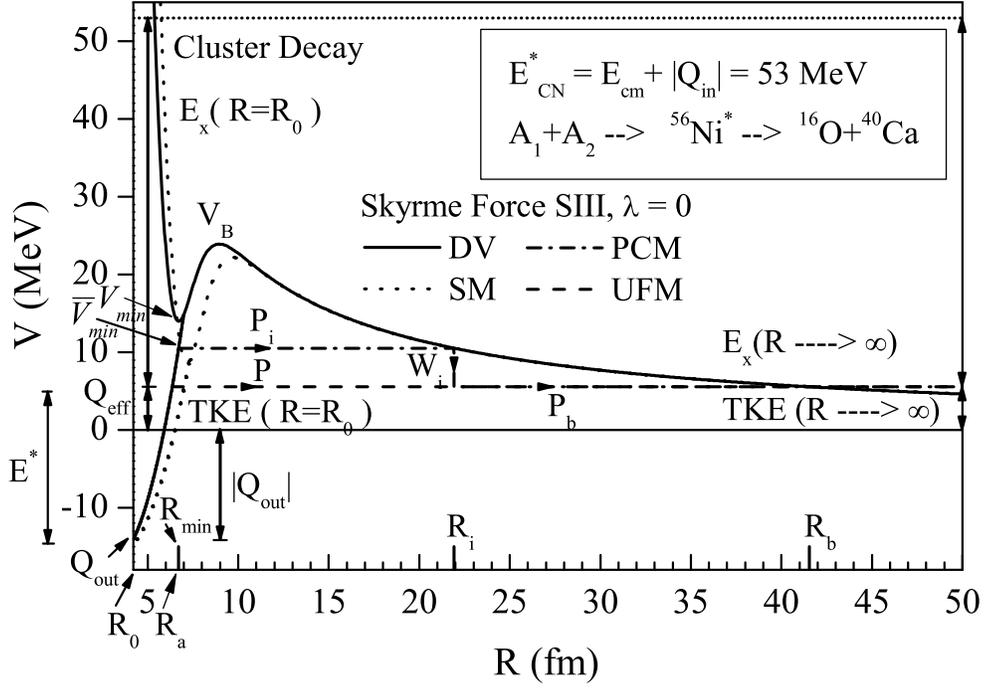}
\vskip -4cm \caption{The scattering potential $V(R)$ (MeV) for
cluster decay of $^{56}$Ni$^{\ast}$ into $^{16}$O + $^{40}$Ca
channel for different Fermi density parameters. The   distribution
of compound nucleus excitation energy E$_{CN}^{*}$ at both the
initial ($R=R_{0}$) and asymptotic ($R \to  \infty$) stages and
$Q$-values are shown. The decay path for both PCM and UFM models
is also displayed.}\label{fig2}
\end{figure}

\begin{figure}[!t]
\centering \vskip 1cm
\includegraphics[angle=0,width=14cm]{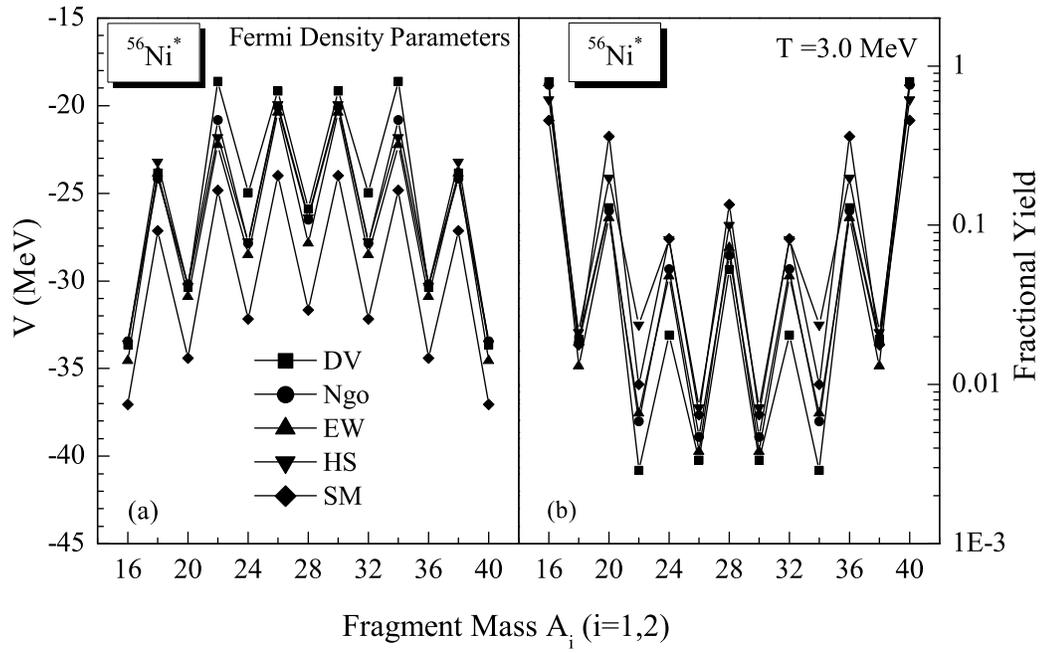}
\vskip -4cm \caption{(a) The fragmentation potential $V(\eta)$ and
(b) calculated fission mass distribution yield with different
density parameters at $T$ = 3.0 MeV.}\label{fig2}
\end{figure}

\begin{figure}[!t]
\centering \vskip 1cm
\includegraphics[angle=0,width=14cm]{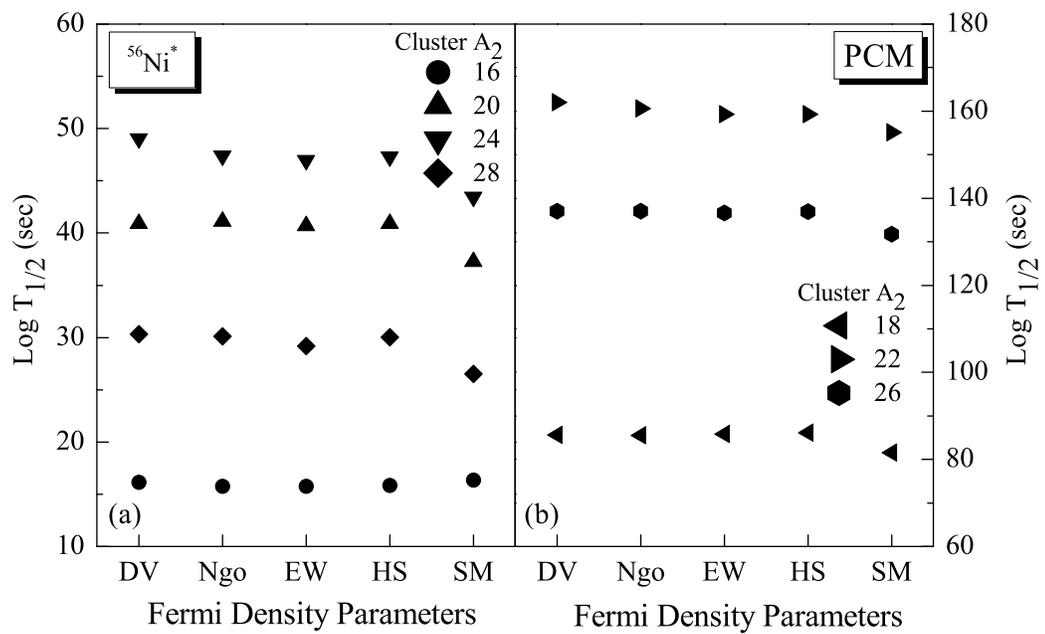}
\vskip -4cm \caption{The variation of $\log  T_{1/2}$ (sec) using
different density parameters for PCM.}\label{fig2}
\end{figure}

\begin{figure}[!t]
\centering \vskip 1cm
\includegraphics[angle=0,width=14cm]{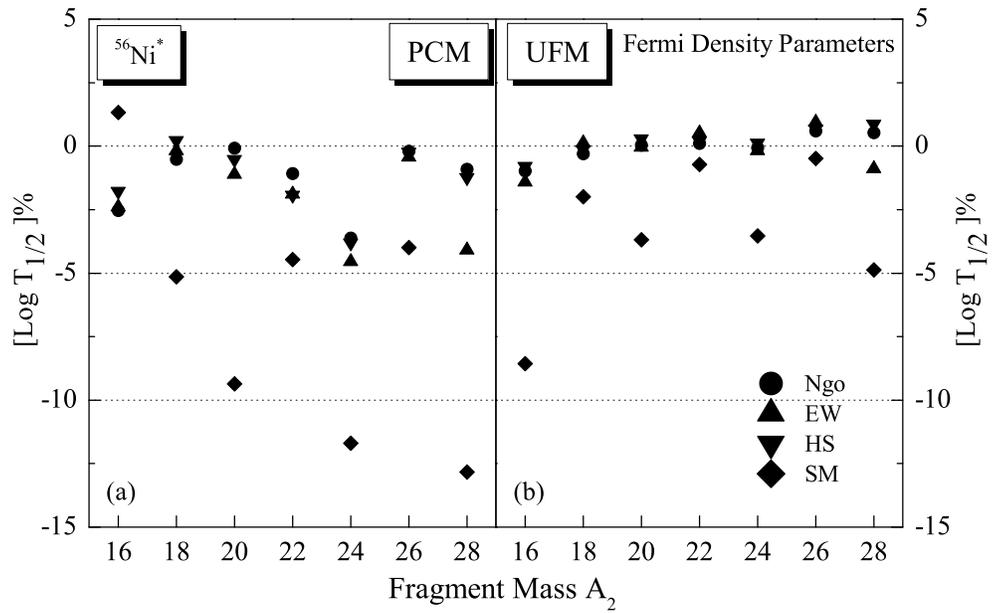}
\vskip -4cm \caption{Percentage variation of $\log  T_{1/2}$ for
different different Fermi density parameters w.r.t. DV
parameters.}\label{fig2}
\end{figure}

\begin{figure}[!t]
\centering \vskip 1cm
\includegraphics[angle=0,width=14cm]{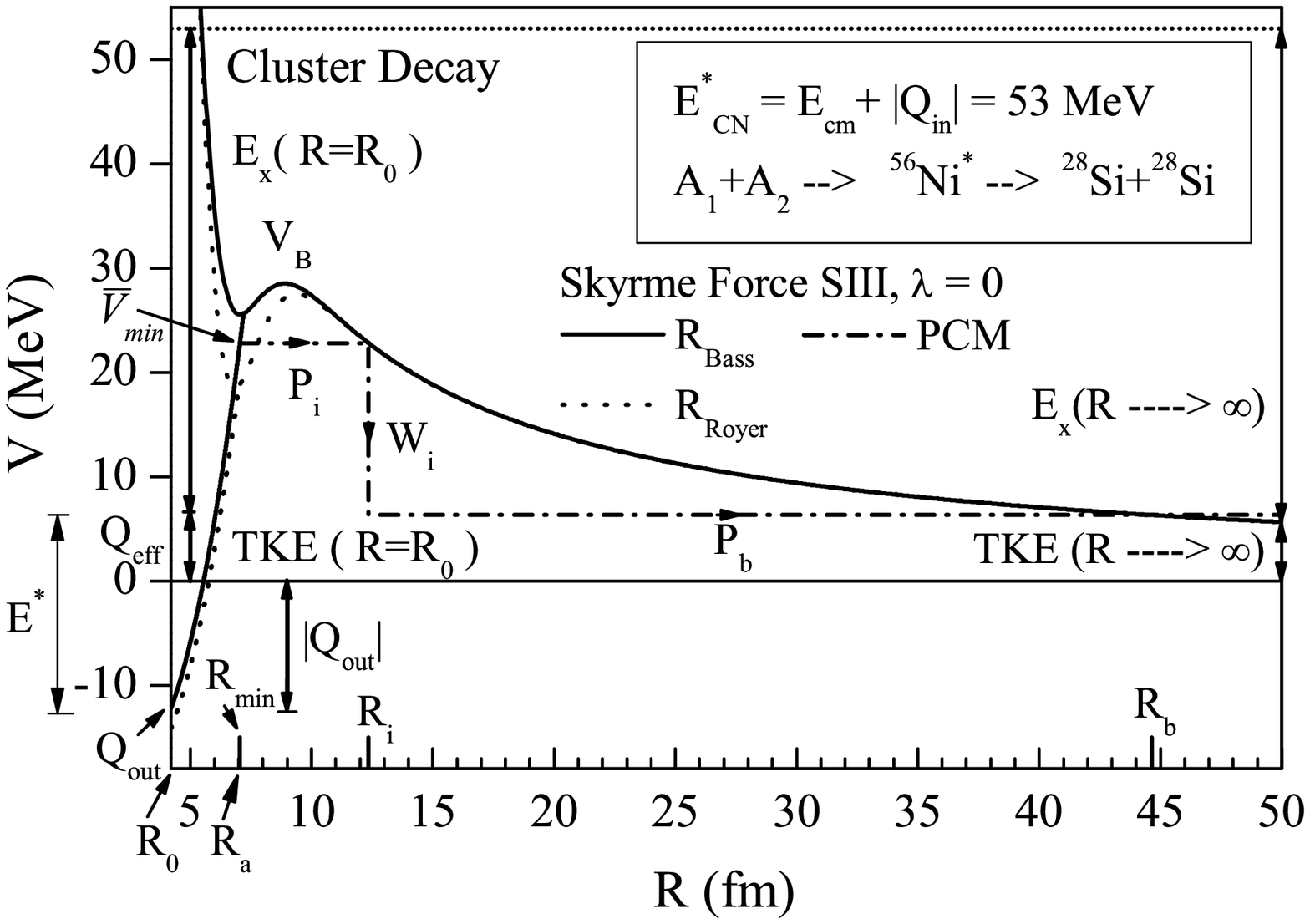}
\vskip -4cm \caption{Same as fig. 2, but for different radii.
 The decay path displayed only for PCM.}\label{fig2}
\end{figure}

\begin{figure}[!t]
\centering \vskip 1cm
\includegraphics[angle=0,width=14cm]{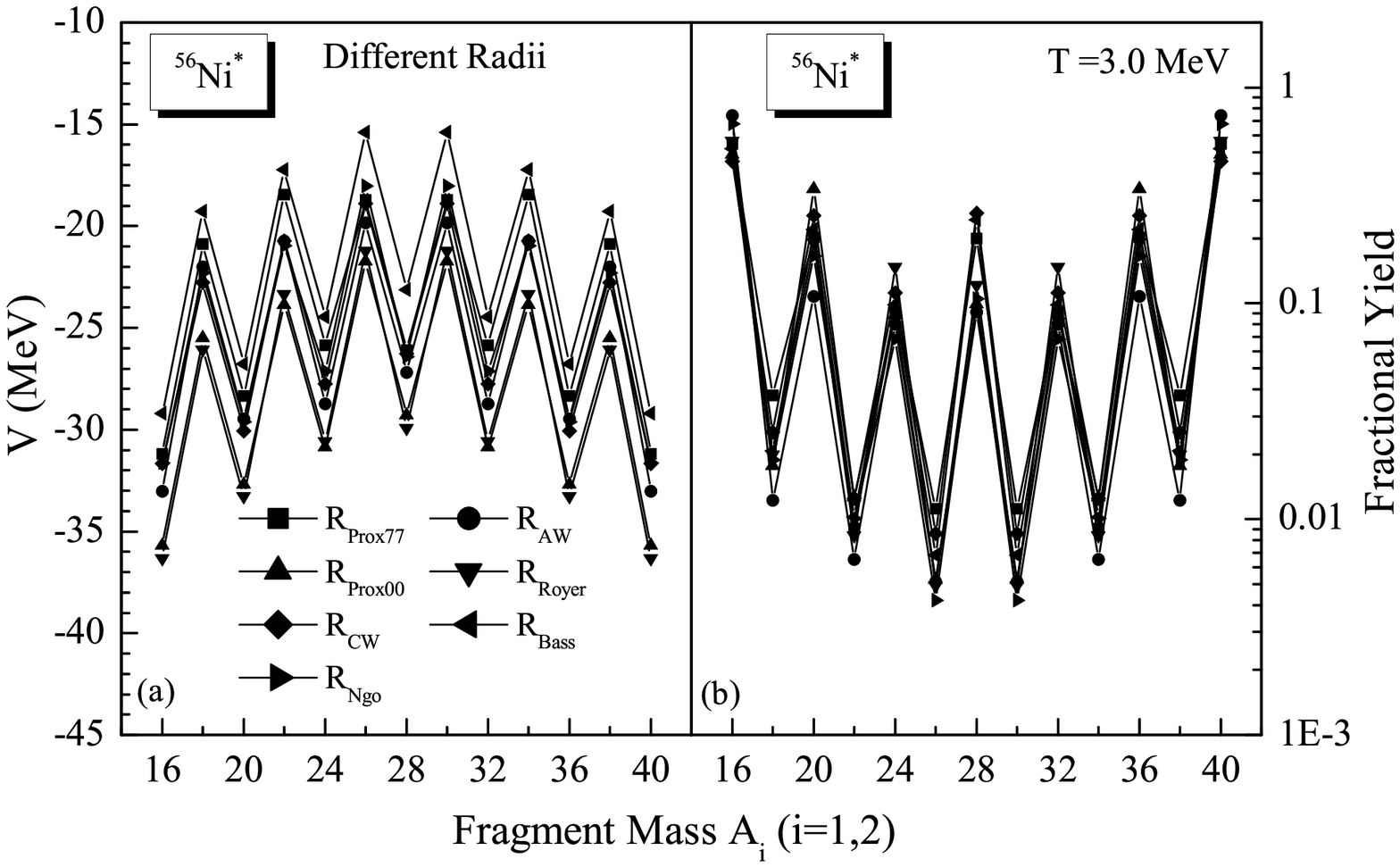}
\vskip -4cm \caption{Same as fig. 3, but for different
radii.}\label{fig2}
\end{figure}

\begin{figure}[!t]
\centering \vskip 1cm
\includegraphics[angle=0,width=14cm]{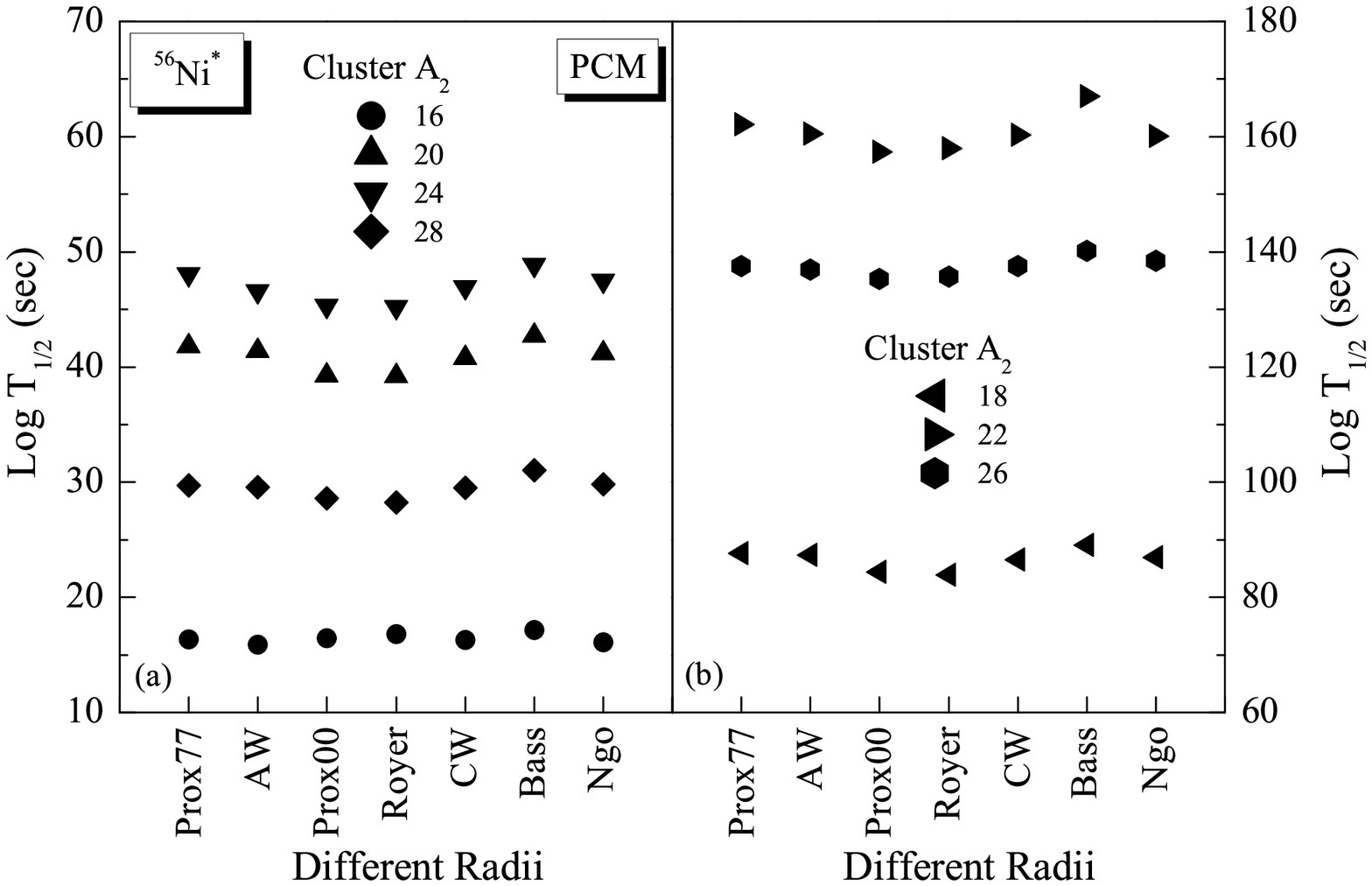}
\vskip -4cm \caption{Same as fig. 4, but for different
radii.}\label{fig2}
\end{figure}

\begin{figure}[!t]
\centering \vskip 1cm
\includegraphics[angle=0,width=14cm]{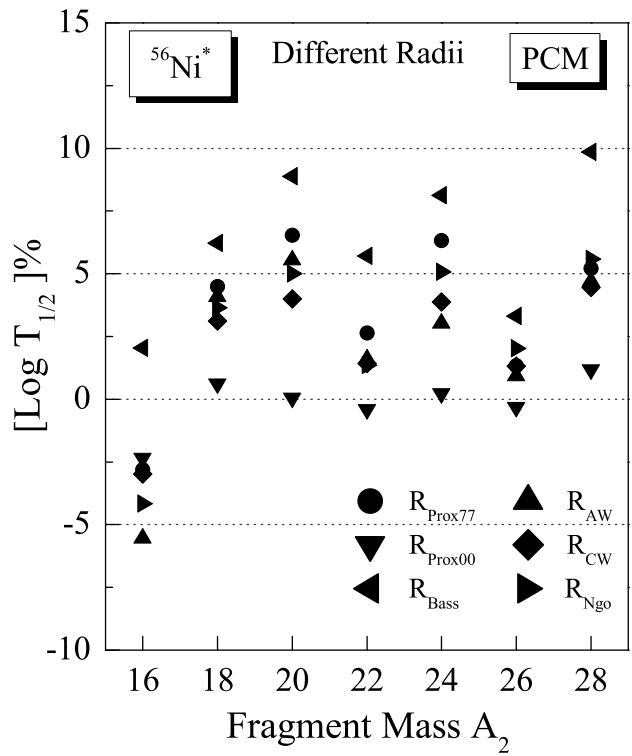}
\vskip -4cm \caption{Percentage variation of $\log  T_{1/2}$ for
different forms of radii with PCM only.}\label{fig2}
\end{figure}

\end{document}